\documentclass[twocolumn,preprintnumbers,amsmath,amssymb]{revtex4-1}
\usepackage{graphicx}
\usepackage{dcolumn}
\usepackage{bm}
\usepackage{color}
\begin{document}

\title{Uniform Shock Waves in Disordered Granular
Matter}

\author{Leopoldo R. G\'omez$^{1}$}
\email{lgomez@uns.edu.ar}
\author{Ari M. Turner$^{2}$}
\author{Vincenzo Vitelli$^{3}$}
\email{vitelli@lorentz.leidenuniv.nl}

\affiliation{
$^1$Department of Physics, Universidad Nacional del Sur - IFISUR - CONICET, 8000 Bah\'ia Blanca, Argentina\\
$^2$ University of Amsterdam, 1090 GL Amsterdam, The
Netherlands\\
$^3$Instituut-Lorentz for Theoretical Physics, Leiden
University, Leiden NL 2333 CA, The Netherlands.}

\date{\today}

\begin{abstract}
The confining pressure $P$ is perhaps the most important parameter
controlling the properties of granular matter. Strongly compressed
granular media are, in many respects, simple solids in which
elastic perturbations travel as ordinary phonons. However, the
speed of sound in granular aggregates continuously decreases as the
confining pressure decreases, completely vanishing at the
jamming-unjamming transition. This anomalous behavior suggests that
the transport of energy at low pressures should not be dominated
by phonons. In this work we use simulations and theory to show how
the response of granular systems becomes increasingly nonlinear as
pressure decreases. In the low pressure regime the elastic energy
is found to be mainly transported through nonlinear waves and
shocks. We numerically characterize the propagation speed, shape,
and stability of these shocks, and model the dependence of the
shock speed on pressure and impact intensity by a simple
analytical approach.
\end{abstract}

\maketitle

\section{Introduction}

Granular systems are materials formed by conglomerates of
macroscopic particles repelling through their mutual contact
\cite{Metha}. These, a priori, simple  systems exhibit complex collective behavior and physical response. For
example, depending on external conditions, granular media are
found in \emph{fluid} or \emph{solid} phases, with defined
characteristic features \cite{JaegerRMP}. Despite the visual
analogy with more conventional matter, the nature of these
granular phases is far from being well understood.

Key to better understanding the behavior and response of
granular systems is the study of the mechanisms of energy
transport and dissipation. Experiments on
elastic wave propagation have shown that the confining pressure
$P$ controls
the speed of sound $c$ of the medium
\cite{Jiaprl1999}-\cite{Jiaprl2004}, causing it to
scale all the way to zero at low pressure. Several works have shown that
for high confining pressures the speed of sound scales as $c \sim
P^{1/6}$, a relation that can be rationalized from an effective
medium theory. However, at low confining pressures a much faster
power law is observed $c \sim P^{1/4}$, whose physical origin
still remains unclear. In general, the complexity of wave
propagation in granular matter is evidenced by the coda-like
signal spectrum of elastic waves, originated from the scattering
of sound with the heterogeneities of the media
\cite{Jiaprl1999},\cite{Jiaprl2004}.

From the theoretical point of view, much of the current understanding of
granular mechanics comes from a simple microscopic approach, where
granular media are modeled as amorphous packings of soft
repulsive spheres \cite{Durian}-\cite{MvHecke}. Using this simple
model, the elastic response has been mainly studied in the linear
regime, by applying a conventional normal mode expansion of the
potential energy \cite{Feng}-\cite{Tighe}. Following this normal
mode approximation, different studies revealed that strongly
compressed granular media have a vibrational spectrum which
follows the Debye law, and therefore, their elastic behavior
parallels the one of simple atomic solids.

Such studies also revealed an anomalous behavior in the elastic
response as the pressure on the medium decreases, and the system
approaches the jamming/unjamming transition. In  this low pressure
regime it was found that the vibrational modes resemble ordinary
phonons only below a characteristic frequency $\omega^*$, that
vanishes as $P$ goes to zero \cite{LSilbert,Wyart}. Above $\omega^*$ the modes are still
extended but strongly scattered by disorder, hence their
ability to transport energy is impaired \cite{Xu,Vitelli}.
This suggests that granular media behave like a continuum elastic medium only
above a characteristic length scale $l^*$ (thought to be related
with $\omega^*$) which diverges as $P$ goes to zero \cite{Wyart}.

However, there is still another source of anomalous behavior in the low
pressure regime, which is universally displayed by granular media,
but it has received far less attention. As a direct consequence of
the nonlinear dependence of the local contact force on the grain
deformations, the sound speed of a granular medium vanishes as $P$
goes to zero \cite{Nesterenko1984}, \cite{NesterenkoBook}. This
fact suggests that the transport of energy in the low pressure
regime should not be dominated by phonons. Actually, linear sound
does not even exist at the jamming point, and a phononic
description here looses its validity. Thus, we can expect that at low
pressures the transport of energy in granular media would be
dominated by supersonic non-linear waves and shocks.

This peculiar property of a vanishing sound speed at non-vanishing
densities, not found in other systems like gases, liquids or
solids, was first studied by V. Nesterenko, who conceived the name
\emph{sonic vacuum} for a media whose sound speed is exactly zero
\cite{Nesterenko1984}, \cite{NesterenkoBook}. Nesterenko and
coworkers mainly analyzed the consequences of the sonic vacuum in
the transport of energy in ordered granular chains
\cite{Nesterenko1984}-\cite{Sen}. In such cases elastic energy was
found to be transported in the form of non-linear solitary waves
(characteristic energy pulses that propagate without shape
distortion).

Despite the results for ordered chains, the mechanisms of
non-linear energy transport in disordered granular matter is largely unexplored \cite{Sinkovits,Gomez}. In particular, it is not clear what role
the structural disorder of the medium plays. Moreover, no systematic study
exist of the expected crossover between the linear waves
obtained at strong confining pressures and the nonlinear waves
expected to dominate in the low pressure regime, when the granular
medium is close to the jamming/unjamming transition.

In this work, we study the non-linear transport of energy in $2D$
disordered granular matter by using numerical simulations and
theoretical calculations. Here we focus on the full response of
two-dimensional granular packings when they are dynamically
compressed by an external piston. These experiments cause the
formation of steadily propagating fronts, without attenuation, nor
deceleration, allowing the clear determination of the nature of
the associated elastic waves (which can be linear, weakly
non-linear, or strongly non-linear). By using this approach, we show
how the elementary excitations of systems at low pressures are
shocks waves rather than ordinary linear phonons \cite{Gomez}. We provide a
full characterization of these shocks, studying features like
shape, propagation speed, and stability. We also present a simple
model which accurately describes the propagation of the shocks,
and the crossover from shocks to linear waves, as the pressure
increases, or the driving strength decreases.

The paper is organized as follow: In section II, we describe the
numerical model and the molecular dynamics simulations. In section
III, we present the numerical results related to the formation of
shock waves by piston compression experiments. In section IV, we
develop an analytical model which captures the speed of the shocks
from the nearly linear to the strongly nonlinear regime using conservation laws.
The appendices present an alternative approach to this behavior:  a continuum
equation of motion (a simplification of Nesterenko's) is derived in
Appendix A while its
soliton and shock solutions are presented in Appendix B.

\section{Numerical model and simulations}

\begin{center}
\begin{figure}[b]
\includegraphics[width=6 cm]{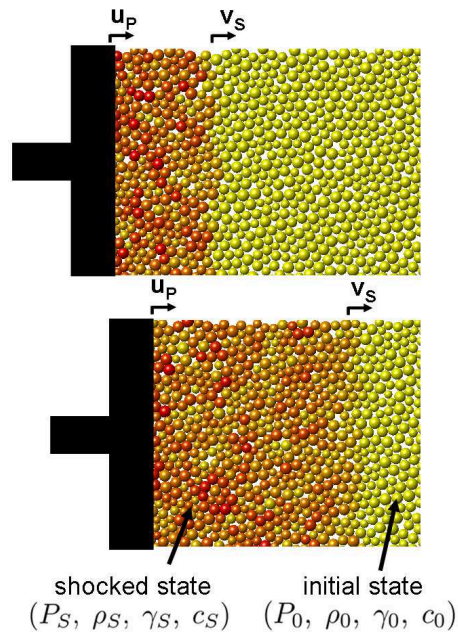} \caption{Piston compression experiments. A
continuous compression of the system by a piston moving at
velocity $u_P$ leads to the formation of a pressure front (shock
wave) travelling at a velocity $v_S$. The shock compression
process can be seen as a fast transition between two states
(initial and shocked) of the granular system.}
\label{fig:schemeshocks}
\end{figure}
\end{center}

Materials with granularity, as diverse as sand, rocks, ceramics,
foams, or emulsion, can exhibit a universal elastic response
which is well described by a simple microscopic approach
\cite{Durian}-\cite{MvHecke}. In all these systems the elasticity
is governed by the soft contact interaction between the grains.
 In
the simplest model of frictionless spherical grains, the contact
potential between two grains $i$ and $j$, located at positions
$\vec{x}_i$ and $\vec{x}_j$, is written as \cite{Somfai}:
\begin{equation}
\begin{cases}
U_{ij}=\frac{\varepsilon_{ij}}{\alpha}\,\,\delta_{ij}\,^{\alpha}
\, \, \,\,\,\,\,\, if \,\,\,\,\, \delta_{ij}>0\\
U_{ij}=0 \, \, \,\,\,\,\,\,\,\,\,\,\,\,\,\,\,\,\,\,\,\,\,\,\,\,if
\,\,\,\,\, \delta_{ij} \leq 0
\end{cases}\label{eq:potential}
\end{equation}
where the overlap between particles is given by $\delta_{ij}\equiv
R_i + R_j-|\vec{x}_i -\vec{x}_j|$, where ${R_i}$ is the radius of
the $i^{th}$ particle. The exponent $\alpha$ gives the strength of the
interaction.  The case $\alpha=5/2$ corresponds to the widely used
Hertz's law. The interaction parameter
$\varepsilon_{ij}=\frac{4}{3}\frac{R_i R_j}{R_i + R_j} E^{*}_{ij}$
is expressed in terms of the effective Young's modulus of the two
particles, $E^{*}_{ij}$; see Ref. \cite{Somfai} for more details.

We model granular media by means of jammed amorphous packings of
particles interacting through the Hertzian law ($\alpha=5/2$). The
packings are prepared at a fixed pressure $P$, or equivalently, an
average particle overlap $\delta_{0} \sim P^{2/3}$, by using a
molecular dynamics algorithm \cite{Somfai}. Similar structures are
also commonly obtained through conjugate gradient algorithms. In
order to avoid crystallization, the particle radii are uniformly
distributed between 0.8 and 1.2 times their average radius. Typical
packings are in the range of $10^3$ to $10^4$ grains with
various width to length ratios.

In this work, we study the transport of energy by means of piston
compression experiments.  Here an initially jammed configuration is
continuously compressed by a piston which moves with a constant
velocity $u_P$ in the $X$ direction, see the schematic in Fig.
~\ref{fig:schemeshocks}. The subsequent motion of the particles is
obtained by numerical integration of Newton's equations of motion
using the Verlet algorithm \cite{Allen}. We employ periodic
boundary conditions in the $Y$ direction and a fixed boundary on
the right edge of the system. In our simulations, the unit of mass
is set by fixing the grain density to unity. The effective
particle Young modulus $E^{*}$ is set to one, which becomes the
pressure unit. These choices ensure that the speed of sound
\emph{inside} the grain, $v_g$ is one \cite{Somfai}. Lengths are
measured in units of average particle diameter.

Piston compression experiments have been previously used in
studies of shock waves travelling through different media, like
gases, liquids, or crystals \cite{Courant}-\cite{MeyersBook}. In
all these systems a travelling pressure front is observed as a
consequence of the motion of the piston. In general, the velocity
and shape of the front reveals important information about the
mechanisms of energy transport.

\section{Numerical Results}

\subsection{Phenomenology of Shocks}

Simulations at different compression velocities show that piston
compression experiments induce the formation of travelling
pressure fronts  (Fig. ~\ref{fig:schemeshocks}). Ahead of the
front there is an undistorted static region where the particles
remain at rest having their initial overlap $\delta_0$. This
initial state is characterized by the original values of pressure
$P_0$, density $\rho_0$, strain $\gamma_0$, and speed of sound
$c_0$. Behind the front we find a compressed region where the
particles have acquired some kinetic energy.

During their propagation the fronts remain essentially flat, such
that to a good approximation the compression experiments can be
described as one dimensional processes. We present a detailed
analysis and discussion of front stability and roughness at the
end of the numerical results.

Taking advantage of the flatness of the front, we average over the
transverse direction $y$, and obtain profiles of the averaged
pressure $P(x)$, density $\rho(x)$, and the $x$ component of the
particle velocity $u(x)$, as a function of the coordinate $x$. Figures 2a
and 2b show the advance of a typical front, as seen through
averaged profiles of pressure and velocity. Note that the velocity
profiles show that behind the shock front the particles move on
average at the piston velocity $u_P$.

\begin{center}
\begin{figure}[t]
\includegraphics[width=8 cm]{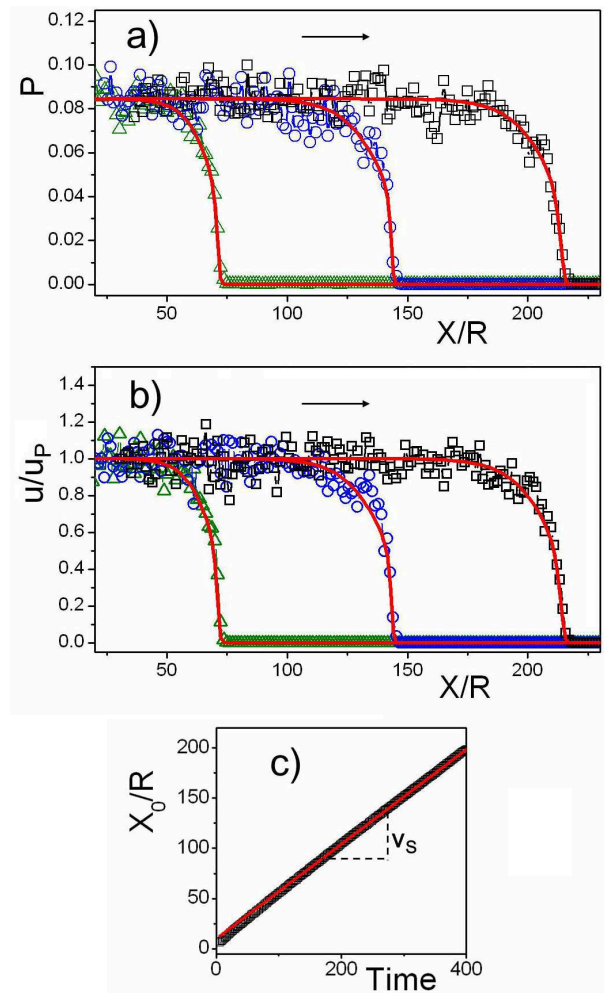} \caption{Phenomelology of shock waves.
Propagation of fronts as seen through the profiles of the averaged
pressure $P$ (a), and averaged particle velocity $u$ normalized
with the piston velocity $u_P$ (b). The front moves in the
direction of the arrows. Different curves corresponds to profiles
at different times and the red lines are phenomenological fits. c)
Position of the front as a function of time. Note that at long
times the front moves linearly with time (red line), with a
propagation speed $v_S$.} \label{fig:shockphenomenology}
\end{figure}
\end{center}

In the steady state regime, the compression front attains a
stationary shape described by an empirical fit, drawn as
continuous lines in Fig. \ref{fig:shockphenomenology}a and
\ref{fig:shockphenomenology}b. In general, the speed of propagation
and the shape of the profiles depend on (1) the degree of
non-linearity of the shocks and (2) the compression
velocity $u_P$. In the next sections, we present a full
characterization of such dependencies. By tracking the position of the
profiles we find that after transients have died out, the front
travels linearly in time $t$, propagating thus with a
characteristic constant speed $v_S$, see Fig. 3. The
determination of the propagation velocity allows the
identification of the compression fronts with non-linear
supersonic shock waves ($v_S \gg c_0$), or linear waves ($v_S=c_0$).

From the averaged profiles it is clear that the particles behind
the front remain in a rather homogeneous state (see Fig.
\ref{fig:shockphenomenology}a and \ref{fig:shockphenomenology}b).
This region is well characterized by the average values
$P_S$, $\rho_S$, $\gamma_S$ and $c_S$. Thus, the passage of the
front can be seen as a transition in the granular media from an
initial ($P_0$, $\rho_0$, $\gamma_0$, $c_0$) to a compressed state
($P_S$, $\rho_S$, $\gamma_S$, $c_S$). This behavior is strikingly
different from the profiles observed in crystals or crystalline
arrays of grains, where large oscillations in pressure, density,
or particle velocity are observed behind the fronts
\cite{NesterenkoBook},\cite{HolianToda},\cite{BringaNuclear}.

\begin{figure}[t]
\begin{center}
\includegraphics[width=8.5 cm]{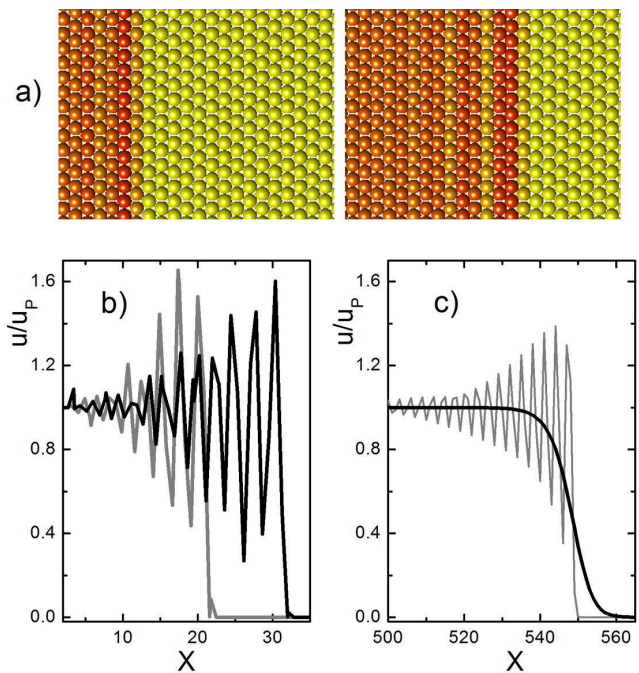} \caption{ Shock waves in a granular
crystal. a) Cartoons of the pressure oscillations as a consequence
of the in-phase motion of crystalline planes (red colors indicate
regions of higher pressure). b) Propagation of the shock wave as
seen through the particle velocity profiles. The front travels
from left to right. c) Shock profiles in the presence of
viscosity, for viscosity coefficients above (thick black line) and
below (thin grey line) critical value. }\label{fig:shockcrystal}
\end{center}
\end{figure}

As an example, Fig. 3a shows the propagation of a shock wave
obtained though the uniform compression of a granular crystal of
hexagonal symmetry with zero pre-compression ($\delta_0=0$).
Figure \ref{fig:shockcrystal}b shows the propagation of the shock
as seen through the particle velocity profiles. Here the large
coherent oscillations are caused by the in-phase motion of the
crystalline planes. Note also that these oscillations change with
time (compare grey and black curves corresponding to the time
evolution), such that the profile never reaches a steady state.

We believe that the underlying mechanism of equilibration
producing the homogenization of the profiles in the amorphous
systems is intrinsically related with the structural disorder. In
our amorphous structures the disorder prevents any coherent
oscillation, such that the peaks are washed out, leaving an
homogeneous state. Similar homogenous states are also commonly
observed in the compression of other unstructured systems, like
gases or liquids \cite{Courant}-\cite{MeyersBook}.

In ordered systems, homogeneous shock profiles are obtained when
some source of dissipation is present
\cite{Duvall},\cite{Herbold}. Figure \ref{fig:shockcrystal}c shows
a steady shock wave profile (black line) obtained when a viscous
dissipation term, based on the relative velocities between
particles, is added to the force: $f_{i,j}=\eta |v_i - v_j|$,
where $\eta$ is the viscous coefficient. In this figure the
viscosity coefficient has been chosen high enough to obtain a
steadily propagating profile without oscillations. For values of
the viscosity coefficient smaller than a critical value, the
obtained fronts reach steady profiles that still display some residual
oscillations (Fig. \ref{fig:shockcrystal}c, gray line). These conclusions, concerning the effect of viscosity
on the shock profile, are demonstrated mathematically in Appendix B.

Although our disordered granular structures do not have any
source of real dissipation, the obtained uniform shock profiles
suggest that the disorder of the system could be playing the role
of an effective viscosity \cite{Gomez}.

\subsection{Jump Conditions}
Mathematically, the propagation of fronts can be studied through
the conservation of mass, momentum, and energy:
\begin{equation}
\begin{cases}
\rho_t+(\rho\,u)_x=0\\
(\rho\,u)_t+(\rho\,u^2+P)_x=0\\
(\frac{1}{2}\rho u^2+U)_t+[u(\frac{1}{2}\rho u^2+U+P)]_x=0
\end{cases}
\end{equation}
In the case of the propagation of uniform shocks, as found in our
simulations (Fig. 2), the (approximate) uniformity of $\rho$, $P$ and $u$ ahead
and behind the shock front allows the immediate integration of
the equations, leading to a set of relationships known as the \emph{
Rankine-Hugoniot jump conditions}
\cite{Courant}-\cite{MeyersBook}:
\begin{equation}
\begin{cases}
\rho_0 v_S=\rho_S(v_S-u_P)\\
\rho_0 \, v_S^2+P_0=\rho_S \, (u_P-v_S)^2+P_S\\
\frac12\rho_0v_S^3+v_S \, U_0 + v_S \, P_0=\\
\frac{1}{2}\rho_S(v_S-u_P)^3+(v_S-u_P) \, U_S+(v_S-u_P) \,
P_S\label{eq:jumpconditions}
\end{cases}
\end{equation}
Here the results are expressed in the reference frame of the shock
front. In this frame, the shock is in a steady state and the net
flow of mass, momentum, and energy toward the shock has to be
zero. The quantities $\rho_0,\rho_S,P_0,P_S,$ and $U_0,U_S$ are respectively
the density, pressure, and internal energy density before and
after the passage of the shock.

In the case of a 1d chain, the density takes the simple form
$\rho=\frac{M}{2R-\delta}$ and the conservation of mass leads to a
relation between the characteristic velocities $u_P$ and $v_S$,
through the average radius of the particles, $R$, and the
compressions in and ahead of the shock, $\delta_S$ and  $\delta_0$ respectively:
\begin{equation}
v_S= u_P\, \frac{2R-\delta_0}{\delta_S-\delta_0} ~. \label{tt}
\end{equation}
Note that since the particle compression $\delta_S$ is typically
much smaller than its diameter $2R$, Eq. (\ref{tt}) implies that
$v_S \gg u_P$. In the limiting case of hard incompressible spheres
$v_S \longrightarrow \infty$. In reality, there is an upper cut-off to the speed $v_s$ given by the sound
speed {\it within} the grain, that is equal in the case of steel to approximately 6000 m/s .

\subsection{Shock's speed}

Commonly, the nature of the traveling fronts (linear or non-linear)
depends on both piston velocity $u_P$ and initial pressure
applied on the system $P_0$. Here, we analyze changes in the
features of wave propagation by tracking variations in the front
speed as a function of piston velocity and the initial confining pressure $P_0$,
which measures the distance of the
uncompressed system to the unjamming transition, which occurs when $P_0 \rightarrow
0$.  This allows us to determine when nonlinear sound propagation takes over
from the linear propagation.

\begin{figure}[b]
\includegraphics[width=8.5 cm]{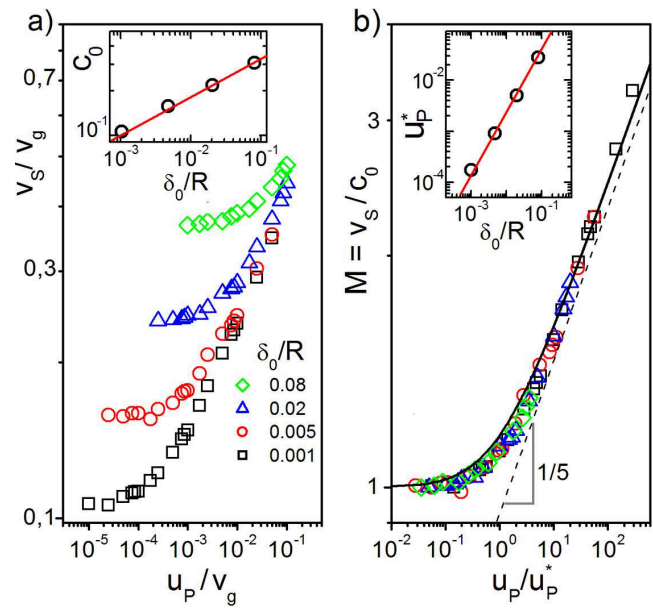} \caption{
Propagation speed. a) Speed of the front $v_S$ versus particle
velocity $u_P$ measured in units of $v_g$, the sound speed within
the grain, for decreasing particle overlap $\delta_0$. The inset
shows the variation of the sound speed $c_0$ with the initial
overlap $\delta_0$ (data) and the scaling $c_0 \sim delta_0^{1/4}$
(line). b) Normalized plot obtained be rescaling $v_S$ with $c_0$
$v_S(0)$ and $u_P$ with $u_P^\star$. The dashed line indicates the
power law $v_S \sim {u_P}^{1/5}$ characteristic of a sonic vacuum.
The inset shows the variation of $u_P^\star$ with the initial
overlap $\delta_0$ (data) and the scaling $u_P^\star \sim
\delta_0^{5/4}$ (line).} \label{frontspeed}
\end{figure}

Figure \ref{frontspeed}a shows the results of this study. Note
that for any finite initial pressure $P_0$, at low compression
velocities $u_P$ the fronts travel with velocities $v_S$ which are
roughly independent of $u_P$. This is evidence of a regime of
linear/weakly nonlinear wave propagation. Here the fronts travel
at approximately the longitudinal speed of sound of the media $v_S
\sim c_0$. In this regime, the propagation speed is simply
controlled by the initial confining pressure $P_0$ or initial
overlap $\delta_0$.

The inset of Fig. \ref{frontspeed}a shows the obtained variation
of the speed of sound $c_0$ with the initial overlap $\delta_0$.
The functional variation of $c_0$ with $\delta_0$ can be
rationalized using scaling arguments. Note that in general $c_0
\sim \sqrt{B}$, where the bulk modulus $B=\frac{\partial
P}{\partial V}$ and $P=\frac{\partial E}{\partial V}$. The change
in volume $dV$ scales linearly with $\delta_0$, the average
overlap between particles, while the energy scales as $E \sim
\delta_{0}^{\alpha}$, see Eq. \ref{eq:potential}. Upon setting
$\alpha=5/2$, we obtain the pressure dependence of the
longitudinal speed of sound $c_0\sim\delta^{1/4}_{0}  \sim
P^{1/6}$ valid for Hertzian interactions \cite{Somfai}. The inset
of Fig. \ref{frontspeed}a shows that the numerical data for $c_0$,
represented by symbols, is consistent with the $\delta_0^{1/4}$
scaling, which is shown as a continuous line.

As the compression velocity $u_P$ is increased, the speed of the
front gradually grows due to a departure from linearity in the
mechanisms of energy transport (Fig. \ref{frontspeed}a). In this
regime elastic energy starts to be mainly propagated through
supersonic shock waves ($v_S>c_0$).

Note that Fig. \ref{frontspeed}a clearly shows that the different
data for $v_S$ vs $u_P$ has the same qualitative structure. This
allows the creation of  a single \emph{master curve}, with the
different data collapsed, as shown in Fig. \ref{frontspeed}b. We
achieve this upon the rescaling of the $v_S$ and $u_P$ axis. The
speed of propagation is normalized with the speed of sound of the
medium $M \equiv v_S / c_0$, where $M$ is the Mach number. The
$u_P$ axis is rescaled by a pressure-dependent velocity scale
$u_{P}^{*}$, which marks the crossover between linear acoustic
waves and shocks.

Above $u_P^*$ , the transport of energy starts to enter into a new
regime dominated by strong shock waves, which travel at two or
three times the speed of sound (Fig. \ref{frontspeed}b). In this
regime we find a power law relation between particle and front
velocities $v_S \sim u_P^{1/5}$ (or $M \sim u_P^{1/5}$) as
indicated by the dashed line.

The main features of this fast compression regime ($u_P \gg
u_P^*$) can also be rationalized through scaling. Since $u_P$, $R$
and $\delta_0$ are all known, we need one additional relation
which combined with Eq.~(\ref{tt}) will make a definite prediction
for the shock speed. We note that for strong shocks, the
propagating front generates a characteristic compression $\delta
\gg \delta_{0}$ and a corresponding increase in the kinetic
energy. In general for strong shock waves the kinetic and
potential energies in the compressed region are of the same order
(\cite{Zeldovich}). In our case this implies $u_{P}^2 \sim
\delta^{5/2}$. We have tested numerically that this non-trivial
proportionality relation exists for strong deformations
\cite{Gomez}. Upon combining the balance between kinetic and
potential energy with Eq.~(\ref{tt}), one readily obtains the
observed power law $v_S \sim u_P^{1/5}$.

\begin{figure}[b]
\begin{center}
\includegraphics[width=8.5 cm]{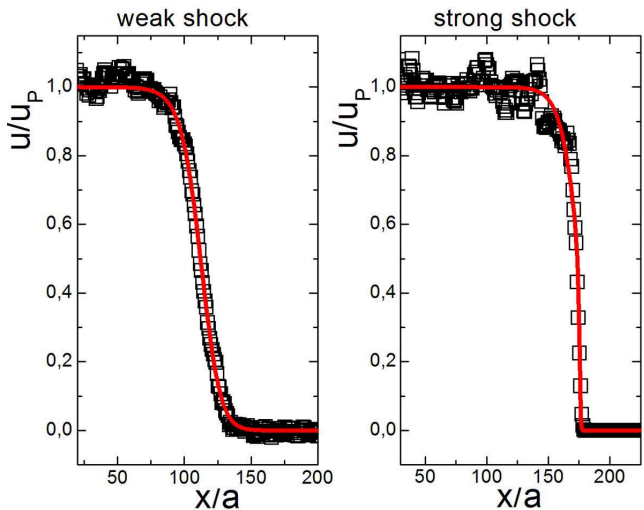} \caption{ Typical shock profiles for
a) weak shock and b) strong shock. The continuous lines represent
phenomenological fits.}\label{fig:shockprofiles}
\end{center}
\end{figure}

We conclude that by controlling $\delta_0$ or $P$, which
parameterize the distance to the jamming point (at $P=0$ and
$\delta_0=0$), we can tune $u_P^*$ and the onset of the strongly
non-linear response of the packings. Our key numerical findings on
the shock velocity summarized in Fig. \ref{frontspeed} can be
grasped from scaling near the jamming point.

\subsection{Shock's shape and width}

Figure \ref{fig:shockprofiles} shows typical shock profiles
observed for weak and strong shock waves. In general, there are
two main significant changes in the profiles due to the increase
in the non-linearity of the phenomena. First, the width of the
shocks decreases when increasing the non-linearity. Second, while
symmetric profiles are obtained for weak shocks, strong shocks are
associated with much more asymmetric profiles. Here we found that
in all these cases, the profile of the shocks can be well
approximated through the phenomenological expression:
\begin{equation}
u(X)=\frac{u_p}{1+\sum_i exp[p_i \, (X-X_0)]}
\end{equation}
where $u_p$ is  the compression velocity, $X_0$ is the position of
the shock wave, and the $p_i's$ are fitting parameters necessary
to describe the profile \cite{Gomez}. While in general weak (symmetric) shock
profiles are well described by only one parameter $p_1$, the
description of  strong shocks requires at least three fitting
parameters $p_1$-$p_3$. The continuous lines of Fig.
\ref{fig:shockprofiles} shows typical fits for weak and
strong shocks.

\begin{figure}[t]
\begin{center}
\includegraphics[width=8.5 cm]{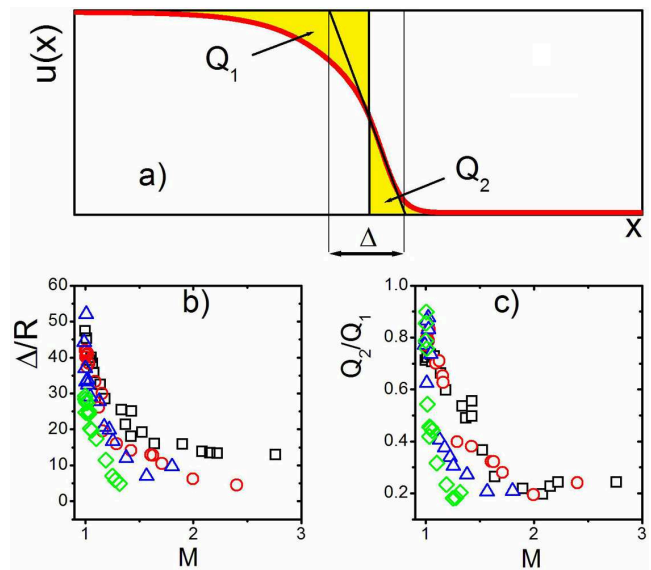} \caption{ Shape characterization of the shock profiles.
a) Scheme of the definition of shock's width and asymmetry.
Variation of shock width (b) and asymmetry (c) with the Mach
number for the different studied shocks. The color code is the
same as in figure \ref{frontspeed}.}\label{fig:shockWQ}
\end{center}
\end{figure}

In order to characterize the features of the shock profiles as a
function of non-linearity, we systematically measure shock widths
and symmetries for the different compression velocities, on
systems with varying initial pressure. We define the shock's width
$\Delta$ and asymmetry parameter $Q$ by the expression
\cite{Grad},\cite{Torrilhon}:
\begin{eqnarray}
\Delta & = & \frac{u_p}{ max(\frac{\partial u}{\partial x})}\\
Q & = & \frac{Q_2}{Q_1}= \frac{\int_{x^*}^\infty u(x) \,
dx}{\int_{-\infty}^{x^*} [u_p -u(x)] \, dx}
\end{eqnarray}
where $x^*$ is such that $u(x^*)=u_p/2$. Figure \ref{fig:shockWQ}a
shows a schematic of the shock profile illustrating the physical
meaning of these definitions.
Figures \ref{fig:shockWQ}b and \ref{fig:shockWQ}c show the results
of shock's width and asymmetry, as a function of the Mach number
of the shocks. The width and symmetry of the shocks monotonically
decrease as the nonlinearity increases. Note that while the width
reaches 50 particles for weak shocks, it reduces to
approximately 10 particles for strong shocks. Also it is clear
that weak shocks are rather symmetric ($Q \sim 1$) and strong
shocks are asymmetric $Q<1$ due to the slow relaxation behind the
front.

\subsection{Characterization of the shocked state}

We have seen how the passage of the pressure front produces a
transition in the granular media from an initial state to a
different shocked state. Since the width of the shock waves is
generally small, the shock compression process is usually
considered to be a fast adiabatic transition
\cite{Courant}-\cite{MeyersBook}; this allows one to trace out the adiabatic
equation of state by measuring the properties of shocks.

\begin{figure}[t]
\begin{center}
\includegraphics[width=8.5 cm]{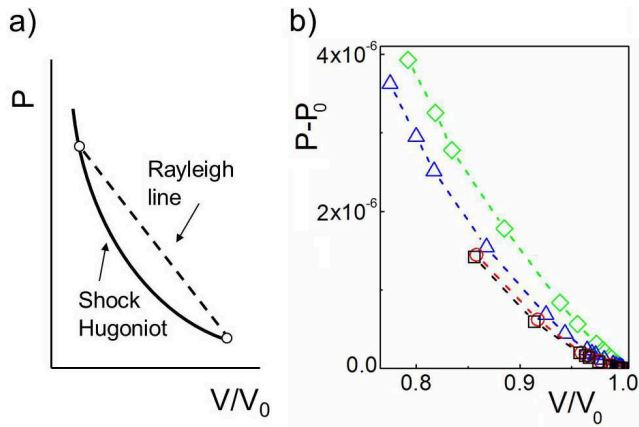} \caption{ a) Schematic of a general shock Hugoniot curve (shock adiabat).
b) Numerically obtained shock Hugoniot for systems with different
initial pressures as obtained from the jump conditions. Dashed
lines are guides to the eyes. \label{fig:shockhugoniot}}
\end{center}
\end{figure}

Generally, the features of the compressed region are studied
through the shock adiabat, or \emph{Hugoniot curve}, which is the
locus of final shocked states. The Hugoniot is an adiabatic line
in the $P-V$ plane (see Fig. \ref{fig:shockhugoniot}a). Each point
in this curve indicates the volume and pressure in the medium in
the final shocked states. Note that during the compression
experiment, which induces the transition from an initial state
$(V_0,P_0)$ before the shock to a final state $(V_S,P_S)$ after
the shock has passed, the system does not
follow the Hugoniot curve in the intermediate states. In general
the evolution is given by the \emph{Rayleigh line} which is the
straight line connecting initial and final states in the $P-V$
plane (see Fig. \ref{fig:shockhugoniot}a).

As the shock compression process is adiabatic, the Hugoniot curve is often
used to construct the equation of state of a given material.
Following this idea, shock compression experiments have been
designed and used for years to study the equation of state of
solids and minerals at high pressures \cite{MeyersBook}.

In our case the shock Hugoniot can be directly obtained from the
simulations (by direct inspection of the final shocked states), or
by using the Rankine-Hugoniot jump conditions. The conservation of
mass and momentum across the shock front Eq.
\ref{eq:jumpconditions}, can be rewritten relating the pressure
and volume in the material with the particle and front velocities
\cite{Courant}-\cite{MeyersBook}:
\begin{equation}
\begin{cases}
V/V_0=1-u_P/v_S\\
P-P_0=\rho_0 \, u_P \, v_S\\
\end{cases}
\end{equation}
When the front velocity is known as a function of particle
velocity $v_S=v_S(u_P)$, these relations become an implicit
expression for the Hugoniot curve. Figure \ref{fig:shockhugoniot}b
shows the shock Hugoniot curve as obtained by using the
Rankine-Hugoniot conditions. In what follows, we use this
numerically determined Hugoniot adiabat to study the stability of
the shock waves for the case of small disturbances in the shape of
the shock front.

\subsection{Shock's stability, roughness, and focusing}

\begin{figure}[b]
\begin{center}
\includegraphics[width=5.5 cm]{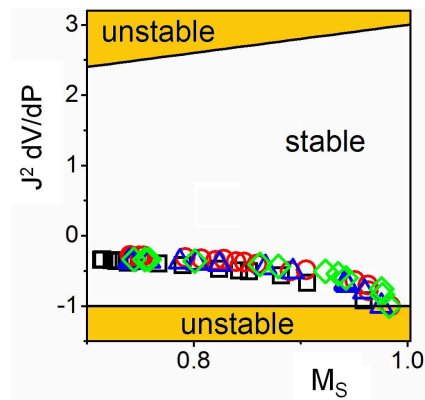} \caption{Dyakov's stability analysis.
Note that all the shock waves belong to the stable region of the
phase diagram.}\label{fig:dyakov}
\end{center}
\end{figure}

The stability of shock waves against shape perturbations has been
one of the interesting and important topics of compressible flow.
The theoretical investigations on this topic started with the
seminal work of Dyakov \cite{Dyakov}, based on a normal mode
expansion of the front perturbations, later re-examined and
presented in more detail by Swan and Fowles \cite{Swan}.

Dyakov performed a linear stability analysis obtaining conditions
for shock stability under which a shock wave remains stable, and
front perturbations decreases exponentially in time. In this
analysis the stability of the shocks is related with the slope of
the Hugoniot adiabat, such that a shock wave remains stable
whenever these conditions hold \cite{Dyakov}, \cite{Swan}:
\begin{equation}
\begin{cases}
j^2 dV/dP < 1 + 2 M_S \\
j^2 dV/dP > -1
\end{cases}
\end{equation}
In these expressions the derivatives are taken along the Hugoniot
adiabat, $M_S$ is the Mach number of the shock respect to the
material behind the front, and $j^2$ is the slope of the Rayleigh
line. In general, when the lower limit of the above equation is
not satisfied the shock wave splits into two waves travelling in
the same direction (splitting instability). The upper limit
corresponds to the corrugation instability, where front
modulations are exponentially amplified.

Note the simplicity and generality of this result. Once the
Hugoniot curve is known these conditions are general and can be
applied to any material. By using our numerically obtained shock Hugoniot, we can perform
the Dyakov test stability analysis. Figure \ref{fig:dyakov}
shows our results, which suggest that the shock
waves studied in this work are stable for the
broad range of compressions and impact intensity analyzed.

Although the shock waves are stable, the disorder of the granular
packings induce the roughening of the front (compare the fronts of
Fig. \ref{fig:schemeshocks} with Fig. \ref{fig:shockcrystal}a). In
general, in crystalline solids shock front's may also become
irregular when travelling through poly-crystalline structures.
Although not a proper instability, changes in the speed of sound
with the lattice orientation can induce the corrugation and
roughening of the shock front \cite{Meyers},\cite{BarberKadau}. In
the amorphous structures considered here, the complete lack of
symmetry produces a rather isotropic speed of sound such that
irregularities in the front are restricted  to a few particles
diameters.

\begin{figure}[t]
\begin{center}
\includegraphics[width=8.5 cm]{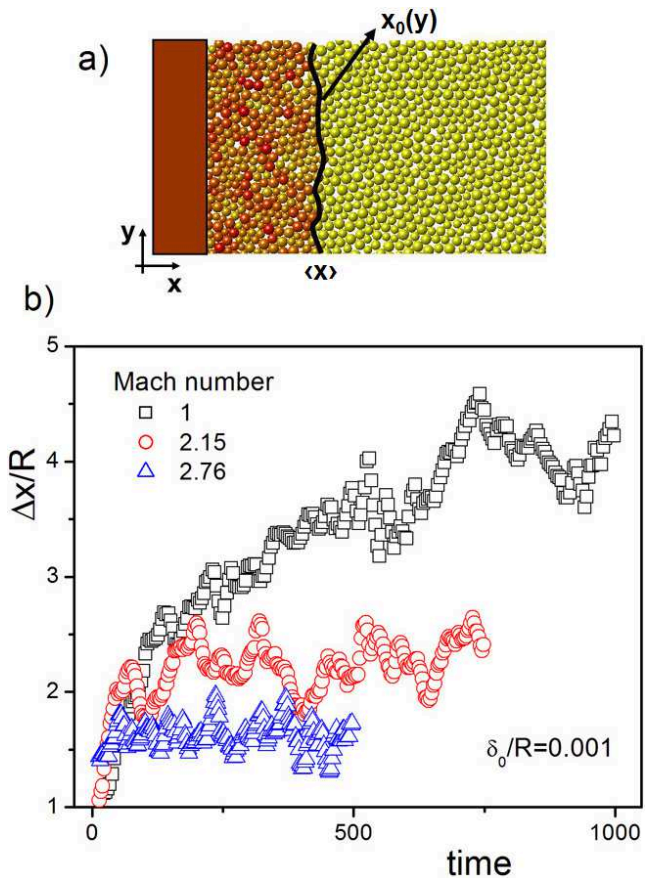} \caption{ Evolution of the front's roughess
as a function of time for different Mach numbers. Note that when
the nonlinearity of the wave is large, the roughness reaches a
plateau at a small value.}\label{fig:roughness}
\end{center}
\end{figure}

In order to analyze variation in front roughnesses we define the
average front position $\langle x \rangle$ and roughness $\Delta
x$ as:
\begin{equation}
\begin{cases}
\langle x \rangle=\langle x_0(y) \rangle\\
\Delta x=\langle (x_0(y)-\langle x \rangle)^2 \rangle^{1/2}
\end{cases}
\end{equation}
where $x_{0}(y)$ is the position of the front at $y$ (see Fig. \ref{fig:roughness}a). Figure
\ref{fig:roughness}b shows the  time evolution of the front
roughness for shocks of varying Mach number. Interestingly,
although the roughness of the fronts seems to reach maximum values
less than $5-6$ particles, the  value of the plateaus clearly
depends on the nonlinearity of the wave. In particular, our simulations suggest
that the more pronounced the nonlinearity of the wave is, the less rough its front is.

\section{Analytical theory of the Shock Speed}

In this section, we analytically describe the relation between the
shock speed and strength, shown in Fig. \ref{frontspeed}, using
the jump conditions Eq. \ref{eq:jumpconditions}. Appendix B gives an
alternative derivation. The first two
equations can be used to predict the speed of the shock if we
assume that the pressure is given entirely by the elastic forces
between the particles. We describe the system with a one-dimensional
one; then any interface cuts only one pair of particles from each
other, and the pressure is just identical to the force between
these two particles, and is given by Hertz's law,
$P=\epsilon\delta^{\alpha-1}$. Combining this equation of state
with the expression for the density $\rho=\frac{M}{2R-\delta}$,
the mass and momentum conservation conditions (first two relations in Eq. \ref{eq:jumpconditions}) imply:
\begin{equation}
\frac{v_S}{c_0}=\sqrt{ \frac{1}{\alpha-1}\,
\frac{(\delta_S/\delta_0)^{\alpha-1}-1}{\delta_S/\delta_0-1}},
\label{eq:shockspeedtheory}
\end{equation}
where $c_0$ represents the speed of sound before the passage of
the shock wave. Together, Eqs. \ref{tt} and
\ref{eq:shockspeedtheory} can be seen as a parametric relation
between front and particle velocities, where the overlap
$\delta_S$ produced by the passage of the front is the parameter.
Such a parametric plot of $v_S$ versus $u_P$ is drawn as a
continuous line on the numerical data in Fig. \ref{frontspeed}b.
This comparison shows that Eqs. \ref{tt} and
\ref{eq:shockspeedtheory} are in excellent agreement with the results of our numerical experiments
on shock propagation.

Despite the good agreement with the numerical results, the third
jump condition is \emph{not} satisfied if we assume the previous
expressions for $v_S$, $u_P$, and energy density per unit length
$U=\epsilon \delta^\alpha\frac{\rho}{M}$ according to Hertz's law.
In the simulations the particles also have a random motion
superimposed on the average velocity $u_P$. This produces an
excess in the pressure and energy of the particles. In this case,
the internal energy has an additional contribution, which
can restore conservation of energy. More precisely, the random
motion is described by an effective temperature $T_S$, and hence
the speed of the shock is no longer overdetermined by the
three jump conditions because the
total number of unknowns, $v_S$, $\delta_S$ and $T_S$ is now equal
to the number of conditions.

These heating effects should change the
shock speed, but not drastically. The prediction above was based entirely on
conservation of mass and momentum, which must still hold. However,
the pressure will have a contribution from the thermal motion, and
this can affect the conclusion. To estimate how big of an effect
the thermal motion has, let us estimate how much energy is
dissipated. First, the velocities may be eliminated from the three
conservation laws to give:
\begin{equation}
\frac{\rho_0U_S-\rho_SU_0}{\rho_S-\rho_0}=\frac{P_0+P_S}{2}
\end{equation}
(this is obtained by solving for the velocities from the first two
conservation laws and substituting into the third). Taking $P_0$,
$P_S$ and $U_0$ to have approximately the Hertzian form (i.e., neglecting
the
temperature),
we can solve for $U_S$. The
excess over the Hertzian potential energy,
$\frac{\rho_S\delta_S^\alpha}{M\alpha}$ is then the thermal
energy, which can be expressed entirely in terms of $\delta_0$ and $\delta_S$:
\begin{equation}
U_{thermal}=\frac{\frac{\varepsilon}{2}(\delta_0^{\alpha-1}+\delta_S^{\alpha-1})(\delta_S-\delta_0)-\frac{\varepsilon}{\alpha}(\delta_S^\alpha-\delta_0^\alpha)}{2R-\delta_S}.
\end{equation}
When $\alpha$ is close to $2$, this energy excess becomes insignificant
since this expression approaches $0$. In particular, when
$\delta_0=0$ and $\alpha=\frac{5}{2}$, $U_{thermal}$ is
$\frac{\alpha-2}{4}=\frac{1}{8}$ of the potential energy,
explaining why our approximation agrees well with the
numerics.

 To summarize, the speed of a shock as a function of the
compression can be estimated from conservation laws if one
neglects the heating of the particles.  In the limit where
$\alpha$ is close to $2$ (as in the Hertzian case) this
approximation is quite accurate.

One could go beyond this approximation and determine the exact relation
between speed and compression in a shock.  First the thermal contribution
to both pressure and energy has to be predicted as a
function of temperature. All the parameters of the shock can then
be determined by solving all three jump conditions simultaneously.

\section{Conclusions}

In this work, we have used simulations of piston compression
experiments to unveil the mechanism of energy transport in
granular systems composed of soft frictionless spheres. While at high confining pressures the elastic
response of the granular system is basically linear, the low pressure
regime is found to be dominated by non-linearities. Thus, for low
confining pressures, when the system is close to the
jamming-unjamming transition, the basic elastic excitations are
supersonic shock waves, rather than linear phonons.

We have presented a detailed characterization of the weakly and
strongly non-linear shocks, studying their propagation speed,
width, shape, and stability. It is interesting to note that the
propagation speed of shocks does not depend on friction or
dissipation. Thus, experiments should focus on this quantity which
is only affected by the contact interaction law. In general, the
presence of friction or dissipation will affect other features
like the width of shocks.

Finally, we developed a mathematical model that correctly captures
the universal nonlinear response of granular packings around the
jamming-unjamming transition. In particular, we have used this
model to describe the dependence of the propagation speed on
pressure and driving.

We conclude by saying that the elastic response of granular matter
present unique features, not found in other condensed matter
systems like solids, liquids, or gases. The athermal character of
these systems allows the attainment of a state with zero pressure at
non-vanishing density, the jamming point, which corresponds to a
sonic vacuum. In this state the response of the system is
characterized by a complete lack of linear behavior. Here we have
shown how this fully nonlinear state also controls the elastic
response of granular system in the low pressure regime.

\textbf{Acknowledgments} We acknowledge helpful discussions with
M. van Hecke, V. Nesterenko, and D. A. Vega. This work was
supported by FOM, Shell, the National Research
Council of Argentina (CONICET), Universidad Nacional del Sur, and
the Instituto de F\'isica del Sur.

\appendix

\section{Equation of motion}

The Lagrangian of a $1d$ chain composed of soft spheres with repulsive interactions, as in Eq. 1, 
reads:
\begin{equation}
L=\sum_n\frac{1}{2}\,m\,\dot{\psi}_n^2-\frac{\varepsilon
}{\alpha}(\psi_n-\psi_{n+1})^\alpha,
\label{eq:spheres}
\end{equation}
where $\psi_n$ is the displacement of the $n$ particle, $m$ is the
mass, $\varepsilon$ the interaction's constant,
and $\alpha$ the interaction's exponent ($\alpha=5/2$ for Hertzian
interactions). The continuum limit of the system is obtained  by
considering the limit where the radius of the spheres $R\rightarrow 0$.
This limit is justified when the wavelength of waves is large compared to the
size of the spheres.  (As Nesterenko found, the wavelength is actually a fixed
multiple of the size of the spheres, but the multiple is large enough that
the continuum approximation is justified.)
Here we proceed to replace
the $\psi_n$'s by a continuous function and perform the continuum
approximation in the Lagrangian, before deriving the equation of
motion.

We first define a continuous function $\phi$ interpolating through
the $\psi_n$'s.  Although the most obvious choice is
$\phi(2Rn)=\psi_n$,  it is not the easiest to implement. In this case
we have $\psi_n-\psi_{n+1}=-2R\phi'(x)
-2R^2\phi''(x)
-\dots$ where $x=2Rn$ is the equilibrium position of the n$^\mathrm{th}$ sphere.
Thus, this approximation will
complicate the calculation of the potential energy, because we
need to use the binomial theorem to raise $\psi_n-\psi_{n+1}$ to
the $\alpha^\mathrm{th}$ power.

In order to get a simpler equation we take the continuum limit
differently. Define $\phi$ by prescribing its derivative instead:
\begin{equation}
\psi_{n+1}-\psi_n=2R\phi'(2R(n+\frac{1}{2})) \,
\label{eq:diff}
\end{equation}
In this case the calculation of the potential energy will be
simple, while the \emph{kinetic} energy will receive corrections from
an expansion in powers of
$R$. (In Nesterenko's version of the equation, dispersion effects appeared
in the potential term so they were nonlinear and more complicated.)
We first invert Eq. (\ref{eq:diff}) in order to
express $\psi_n$ in terms of $\phi$. To do that we make a Taylor
expansion of the left-hand- side about $n+\frac{1}{2}$.  This
gives (up to terms of order $(2R)^4$):
\begin{equation}
\psi'(2R(n+\frac{1}{2})) +\frac{(2R)^2}{24}
\psi'''(2R(n+\frac{1}{2}))=\phi'(2R(n+\frac{1}{2})) .
\end{equation}
Integration of both sides with respect to $x$ leads to
$\phi(x)=(1+\frac{R^2}{6}\frac{d^2}{dx^2})\psi(x)$.
Solving for $\psi$ in terms of $\phi$ gives:
\begin{equation}
\psi(x)=\frac{1}{1+\frac{R^2}{6}\frac{d^2}{dx^2}}\phi(x).
\end{equation}
Expanding the ratio in a geometric series leads to
$\psi(x)\approx
\phi(x)-\frac{R^2}{6}\phi''(x)\dots$. Thus
the kinetic energy term can be approximated by:
\begin{equation}
\frac12\dot{\psi}_n^2\approx\frac12\dot\phi(x)^2-\frac{R^2}{6}\dot{\phi}(n)\dot{\phi}''(x)
\end{equation}

The approximated continuum Lagrangian is obtained by replacing the
$\psi$ in terms of $\phi$ and the sum by an integral:
\begin{eqnarray}
L\approx\int
\left(\frac{1}{2}\dot{\phi}^2-\frac{R^2}{6}\dot\phi\dot\phi''
-\frac{\varepsilon (2R)^\alpha}{m \alpha}(-\phi')^\alpha\right) \frac{dx}{2R}\nonumber\\
\approx\int
\left(\frac{1}{2}\dot{\phi}^2-\frac{R^2}{6}(\dot\phi')^2
-\frac{\varepsilon (2R)^\alpha}{m \alpha}(-\phi')^\alpha\right) \frac{dx}{2R}.\nonumber\\
\label{eq:take3}
\end{eqnarray}
Finally, by using the Euler-Lagrange equation, we obtain the
equation of motion:
\begin{equation}
\ddot{\phi}-\frac{R^2}{3}\,\ddot{\phi}''+\frac{\varepsilon (2R)^\alpha}{m}
[(-\phi')^{\alpha-1}]\,'=0 \label{motionphi}
\end{equation}
In terms of the compression field $\delta=-2R\phi'$ this equation
takes the form:

\begin{equation}
\ddot{\delta}-\frac{R^2}{3}
\,\ddot{\delta}''-\frac{4R^2\varepsilon}{m} \,
[\delta^{\alpha-1}]\,''=0
\end{equation}

This equation is simpler to the equation originally derived by
Nesterenko; there is more than one way to take the continuum limit
when the wavelength is the same order as the particle spacing.
Here we found
that this simpler equation is enough to describe the simulation
results.

\section{Steady state solutions}

The travelling of stationary perturbations having a propagation
velocity $v_S$ can be considered by replacing the ansatz
$\delta(x,t)=\delta(\tilde{t}\equiv t-\frac{x}{v_S})$ in the equation of
motion.  This function $\delta(\tilde{t})$
describes what an observer at a fixed point of space
sees as the shock passes by. It satisfies the ordinary differential equation:
\begin{equation}
\frac{R^2}{3v_S^2}\,\frac{d^4\delta}{d\tilde{t}^4}-v_S^2
\frac{d\delta}{d\tilde{t}}+\frac{\varepsilon}{mv_S^2}\frac{d^2[\delta^{\alpha-1}]}{d\tilde{t}^2}=0\nonumber
\end{equation}
Integrating twice and applying the boundary conditions
$\delta(\tilde{t}\rightarrow -\infty)=\delta_0$,
$\dot\delta(\tilde{t}\rightarrow -\infty)=\ddot\delta(\tilde{t}\rightarrow-\infty)=0$,
leads to the following expression:
\begin{equation}
\frac{R^2}{3v_S^2}\,\ddot\delta-
\delta+\frac{4\varepsilon R^2}{mv_S^2}\delta^{\alpha-1}=
-\delta_0+\frac{4\varepsilon R^2}{mv_S^2}\delta_0^{\alpha-1}\nonumber
\end{equation}
This last equation can be rewritten in the form:
\begin{equation}
\frac{d^2\delta}{d\tilde{t}^2}=-\frac{dW}{d\delta},
\end{equation}
where the field $W$ is given by:
\begin{equation}
\begin{split}
W(\delta,v_S)&=\frac{12 \varepsilon }{m\alpha}\,(\delta^\alpha-\delta_0^\alpha)-\frac{3v_S^2}{2R^2}(\delta-\delta_0)^2
\\&-\frac{12\varepsilon}{m}\delta_0^{\alpha-1}(\delta-\delta_0).
\end{split}\label{eq:field}
\end{equation}
Thus, the problem of traveling steady waves is mapped to
the motion of an \emph{effective} particle moving (in time
$\tilde{t}$) in a potential field $W$. Note that in this case the
potential not only depends on the compression $\delta$, but also
on the propagation speed $v_S$.

\begin{center}
\begin{figure}[t]
\includegraphics[width=6.5 cm]{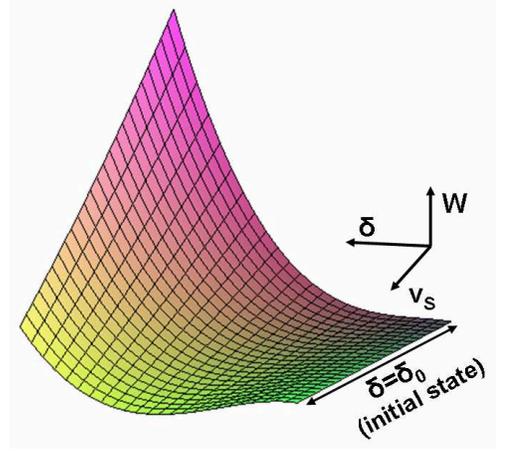} \caption{Effective potential surface.
For $v_S>c$ the initial state becomes a local maxima allowing the
description of travelling perturbations.} \label{potentialsurf}
\end{figure}
\end{center}

We now use this analogy to study the properties of travelling
waves. Figure \ref{potentialsurf} shows the shape of the potential
as a function of compression and propagation speed. In the initial
state, the system is undistorted ($\delta=\delta_0$) and the
effective particle is located at the extreme $W(\delta_0,v_s)=0$,
at rest. For the cases where this initial state is a local minima,
the effective particle remains at this initial position for all
times. This means that in these cases steady perturbations cannot
travel through the system. Thus, in order to describe travelling
perturbations the initial state needs to be a local maximum such
that $W_{\delta \delta}(\delta_0,v_S)<0$, and the speed of
propagation needs to satisfy:
\begin{equation}
v_S^2\geq\frac{4R^2(\alpha-1)\varepsilon}{m} \delta_0^{\alpha-2}
\end{equation}
Physically, this condition means that travelling perturbations
will propagate at the speed of sound of the medium
$c_0=2R\sqrt{(\alpha-1)\varepsilon/m}\,
\delta_0^{\frac{\alpha-2}{2}}$, or faster. Figure
\ref{potentialsurf} shows the change in the nature of the initial state,
from a local minima to a local maxima, as the speed of propagation
of the waves is increased.

\subsection{Solitary waves}
\begin{center}
\begin{figure}[t]
\includegraphics[width=7.5 cm]{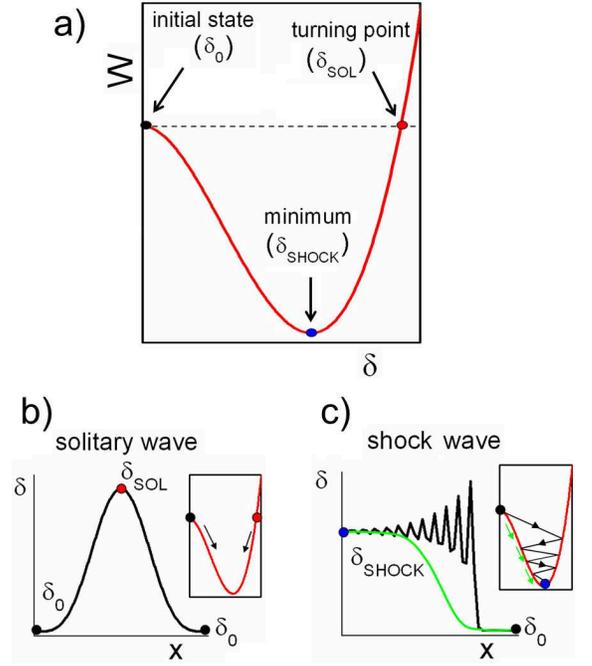} \caption{Steady state solutions
of the non-linear wave equation. a) Potential well for a given
propagation speed $v_S$. b) Solitary wave solutions obtained in te
absence of viscosity. c) Shock wave solutions obtained for viscous
dissipation.} \label{solitonsshocks}
\end{figure}
\end{center}

Now consider a travelling perturbation which satisfies the above
condition ($v_S \geq c_0$). Figure \ref{solitonsshocks}a shows the
shape of the potential $W(\delta)$ for a given front velocity
$v_S$. Far away, the undistorted system corresponds to the
effective particle located at the maxima of the potential $W=0$.
The distortion induced by the passage of the wave can be seen as
the particle rolling down the potential. The effective particle
will move up to the turning point, and then the conservation of
energy implies that the particle would roll back to the initial
state.

This motion of the effective particle corresponds to a solitary
wave travelling trough the system with velocity $v_S$ (Fig.
\ref{solitonsshocks}b). Note that these solitary waves are the
only steady state solutions of the system. The speed of
propagation can be obtained from the turning point condition
$W(\delta_{SOL},v_S)=0$, where $\delta_{SOL}$ is the maximum
compression due to the passage of the wave:
\begin{equation}
\frac{v_S}{c_0}=\sqrt{\frac{2}{\alpha(\alpha-1)}
\left[\frac{\delta_{SOL}^\alpha-\delta_0^\alpha}{\delta_0^{\alpha-2}(\delta_{SOL}-\delta_0)^2}
-\frac{\alpha\delta_0}{\delta_{SOL}-\delta_0}\right]}
\end{equation}

\subsection{Shock waves} The previous analysis shows that in the absence of any source of dissipation the
motion of the effective particle conserves its energy, and the
only steadily propagating solutions are solitary waves. In that
case homogeneous shocks, as found through the simulations in this
work, are not possible solutions of the equation of motion.

In order to obtain steady state shock solutions, some sort of
dissipation is needed. In general, the effect of dissipation can
be taken into account by adding new terms in equation
\ref{motionphi}. For example, if a dissipation term (due to relative
motion of the spheres) is
considered \cite{Herbold}, the equation of motion of the system
will take the form:
\begin{equation}
\ddot{\phi}-\frac{R^2}{3}\,\ddot{\phi}''+\frac{4R^2\varepsilon}{m}
[(-\phi')^{\alpha-1}]\,'-4R^2\eta\dot{\phi}''=0,
\end{equation}
where $\eta$ is a damping coefficient.

Now if we look for steady state travelling perturbations with the
same procedure as before, we obtain the following equation:
\begin{equation}
\frac{d^2\delta}{d\tilde{t}^2}=-\frac{dW}{d\delta}-12\eta\frac{d\delta}{d\tilde{t}}
\end{equation}
Here the potential $W$ is the same as obtained before
(equation (\ref{eq:field})). Thus, for this particular form of dissipation, we
have explicitly shown that the problem of steadily propagating
waves is mapped to the problem of an effective particle moving in
a potential, but now the particle also moves under the effect of
friction.

In this case, the presence of friction will take out the energy
of the effective particle, inducing the oscillation to decay to the minimum
of the potential well. These solutions correspond to
oscillatory shock waves (Fig. \ref{solitonsshocks}c, black line).

If the viscosity is high enough, the motion of
the effective particle will become overdamped, and the particle
directly moves to the minimum of the potential, without performing
any oscillation. This motion corresponds to
traveling regular shock profiles (Fig.
\ref{solitonsshocks}c, green line).

For the shock waves the relation between propagation velocity and
induced compression can be obtained by the condition
$\frac{dW}{d\delta}_{\mid \,\delta_{S},v_S}=0$, which corresponds
to the minimum of the potential. This
condition leads to:
\begin{equation}
\frac{v_S}{c_0}=\sqrt{ \frac{1}{\alpha-1}\,
\frac{(\delta_S/\delta_0)^{\alpha-1}-1}{\delta_S/\delta_0-1}},
\end{equation}
where $\delta_S$ is the maximum compression after the shock
wave has passed.
In Sec. IV, imposing two of the jump conditions (conservation of momentum and mass, but not energy) produced the same result. The dissipation term we have added
preserves the same two conservation laws, explaining
the agreement.  In Sec. IV, we justified
the disappearance of energy not by introducing friction (since
the simulations did not include it) but by assuming
that the energy
is converted to random motion of the particles. A more accurate
treatment of this hypothesis would have to account for some additional thermal pressure.


\end{document}